\documentclass[12pt]{article}
         
\usepackage{amsmath,amssymb} 
\usepackage{pictfile}     
\usepackage{latexsym,amsfonts,amstext,verbatim}            
\usepackage[dvips]{graphicx}

\numberwithin{equation}{section}

\renewcommand{\[}{\hspace{0.0125em}\relax}

\begin{document}

 \title{Wave-particle duality:\\ suggestion for an experiment}

\author{ {R N Sen}\\
{\normalsize Department of Mathematics}\\[-0.7mm]
{\normalsize Ben-Gurion University of the Negev}\\[-0.7mm] 
{\normalsize 84105 Beer Sheva, Israel}\\
{\normalsize\tt E-mail: rsen@cs.bgu.ac.il}}


\maketitle\thispagestyle{empty}

\begin{abstract}

\thispagestyle{empty}

\vspace{2mm}
\noindent
Feynman contended that the double-slit experiment contained the `only
mystery' in quantum mechanics.  The mystery was that electrons
traverse the interferometer as waves, but are detected as particles.
This note was motivated by the question whether single electrons can
be detected as waves. It suggests a double-slit interferometry
experiment with atoms of noble gases in which it may be possible to
detect an individual atom as a probability wave, using a detector
which can execute two different types of simple harmonic motion: as a
simple pendulum, and as a torsion pendulum. In the experiment, a
torsional oscillation will never be induced by the impact of a
probability wave, but will always be induced by the impact of a
particle. Detection as a wave is contingent on the atom interacting
much more strongly with the macroscopic detector \emph{as a whole}
than with its microscopic constituents. This requirement may be more
difficult to meet with electrons, protons, neutrons or photons than
with atoms. 

\end{abstract}
\pagebreak



In his famous \emph{Lectures} \cite{FLS2006}, Feynman stated that
double slit experiments contain the `only mystery' in quantum
mechanics. By `mystery' he apparently meant phenomena that could not be
understood in terms of classical physics. This failure of classical
physics could be traced to wave-particle duality, and wave-particle
duality could be illustrated simply, and convincingly, by the
double-slit experiment with electrons.

In Feynman's gedankenexperiment, electrons were detected \emph{as
particles} (by a geiger counter or electron multiplier), but
traversed the interferometer as \emph{waves}. If one tried to detect
which slit an electron went through, the interference pattern was
lost. Feynman traced this loss to the position-momentum uncertainty
relation.
   
The passage of a microscopic object through an
interferometer\footnote{Owing to the centrality of the detector to
our considerations, we shall restrict use of the term
\emph{interferometer} to mean only the system of slits or gratings.
\label{INTER}} may be regarded as an \emph{interaction} which changes
the state of the object but not that of the (macroscopic)
device.\footnote{Note the similarity with external field problems in
quantum electrodynamics \cite{JR1955}.} Feynman's double-slit
experiment may then be viewed as a succession of two interactions:
(i)~the electron-interferometer interaction, and (ii)~the
electron-detector interaction. In (i), the electron acts like a wave;
in (ii), like a particle. The question which does not seem to have
been asked is: can (ii) be replaced by an interaction in which the
electron, or some other microscopic object, acts like a wave?  The
answer to this question may be in the affirmative for suitable pairs
of microscopic objects and detectors, and the aim of this note is to
suggest a double-slit experiment with \emph{atoms} for testing this
possibility.

\begin{figure}[ht]
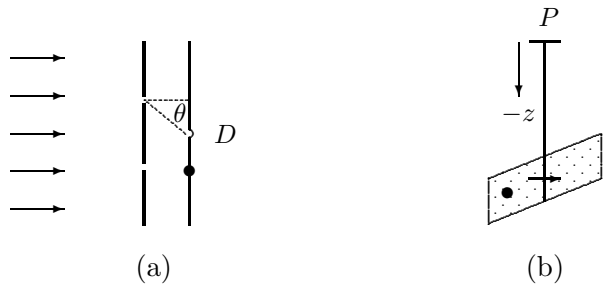


\footnotesize
\beginpicture
\small
\setcoordinatesystem units <0.5mm,1mm>

\setplotarea x from -35 to 45, y from -20 to 19

\linethickness=1pt


\putrule from 61 -12.3 to 61 -5  %
\putrule from 61 -4    to 61 4      %
\putrule from 61 5     to 61 12.3    %


\linethickness=0.4pt

\putrule from 73 -12.3 to 73 -0.5
\putrule from 73 0.5 to 73 12.3

\circulararc 180 degrees from 73 -0.5  center at 73 0


{\setdashpattern <1pt,1pt>

\plot 72 0  61 4.5  73 4.5 /}

\put {\footnotesize$\theta$} at 68 2.75


\linethickness=0.6pt

\multiput {$\vector(1,0){20}$} [B1] at 30 -10   30 -5  30 0  
30 5  30 10 /


\linethickness=0.4pt

\plot 150 -12  180 -6  180 0  150 -6  150 -12 /

\setshadegrid span <3pt>

\vshade 150 -12 -6  180 -6 0 /


\linethickness=.7pt

\putrule from 165 -9 to 165 12.3

\putrule from 161 12.3 to 169 12.3

\put {$\vector(1,0){12}$} [B1] at 165 -6


\put {$\vector(0,-1){20}$} [B1] at 158 12

\put {\footnotesize $-z$} at 158 2.5


\multiput {$\bullet$} at 155 -8   70.5 -5 /


\put {\footnotesize$D$}  at 80 0
\put {\footnotesize$P$}  at 166 15.5

\put {(a)} at  61 -18
\put {(b)} at 165 -18

\endpicture
\caption{Testing whether atoms arrive as particles or waves}
\label{FIG}
\end{figure}

Figure \ref{FIG} illustrates the scheme of the experiment and the
design of the detector. The left-hand side (a) shows a cross-section
of the scheme in the (horizontal) $xy$-plane. Here the $z$-axis is
normal to the plane of the paper. The right-hand side (b) shows the
detector $D$, which is a thin rectangular vane suspended from the
top. Here the plane of the paper is the $xz$-plane and the $y$-axis
is normal to it. The key point is that the vane must be able to
execute two essentially different types of simple harmonic
motion:\footnote{We are assuming that the only motion that the vane
can execute relative to the line of suspension is torsional
oscillation.  This is an assumption on the coupling between the vane
and the suspension.} (i)~as a simple pendulum in a plane through the
$z$-axis, and (ii)~as a torsion pendulum \emph{around} the $z$-axis,
the line of suspension of the pendulum at rest.\footnote{The idea of
using such a device was inspired by the \emph{Nichols radiometer},
the torsion balance used by Nichols and Hull \cite{NH1901} to
demonstrate the pressure of electromagnetic radiation in 1901. The
effect was observed independently be Lebedev \cite{PL1901} in the
same year.} The vane is set in motion by the impact of an atom.  The
atoms are sent through the interferometer \emph{one at a time}.

Assume, now, that interaction between the atom and the detector
transfers momentum to the detector \emph{but has no other effect upon
it}; the vane, which was initially at rest, is set in motion.  The
experiment consists of repeated observation of the motion of the vane
caused by the impact of a single atom. After observing the motion
caused by the $n$-th impact, the vane has to be brought to rest
before the $(n+1)$-th impact; \emph{the experiment is not designed to
reveal the interference pattern}.

As one sees form Fig.~\ref{FIG}(a,\,b), the two slits are
symmetrically placed with respect to the central vertical axis of the
vane.  Therefore the single-particle probability distribution
$|\psi|^2$ at the vane will be symmetric with respect to this axis.
If the vane `sees' $|\psi|^2$ much as the human eye sees ripples on
the surface of a pond, then the \emph{momentum transfer to the vane
ought to be symmetric about its central axis} (along the little arrow
through the vane shown in Fig.~\ref{FIG}(b)). The vane will then be
set in motion as a \emph{simple} pendulum, in the $xz$-plane of
{Fig.}~\ref{FIG}(b). 

If, on the other hand, the vane `sees' the atom as a particle
(striking at the point marked by bullets in Fig.~\ref{FIG}), the atom
will impart to it: (i)~a \emph{torque} around the $z$-axis; and
(ii)~linear momentum, along the plane of the vane (see
Fig.~\ref{FIG}(b)). The resulting motion of the vane will be a
superposition of two motions: (i)~oscillation as a \emph{torsion
pendulum} around the $z$-axis; and (ii)~motion as a simple pendulum
in the plane of the vane (the $yz$-plane).

The above can be rephrased (somewhat loosely) as follows. In the
experiment described above, \emph{the detector will respond
differently to waves and to particles}.  If it is struck
symmetrically by a wave, it will \emph{never} execute torsional
oscillations; if struck by a particle, its motion will \emph{always}
have a component of torsional oscillation. The angular frequency of a
torsion pendulum is $\omega_{\sf t} = \sqrt{k/I}$, where $k$ is the
torque constant of the suspension wire and $I$ the moment of inertia
of the vane around its axis of suspension. The angular frequency of a
simple pendulum is $\omega_{\sf s} = \sqrt{g/L}$, where $L$ is the
length of the suspension. That is, $\omega_{\sf s}/\omega_{\sf t} =
\sqrt{gI/kL}$. The experimenter has considerable control over the
parameters $I$ and $L$. By varying them, it should be possible to
control the relative sensitivity of the detector to waves and to
particles.

If the vane is struck by a {particle} (from one of the slits) close
to normal incidence and sufficiently close to its centre, the motion
that results may be indistinguishable, within experimental error,
from that resulting from the impact of a wave. This contingency will
not arise if the angle $\theta$ is large enough, which will be
assured if the distance between the plane of the slits and the
detector at rest is not nuch larger than the distance between the
slits, as shown in the figure. Furthermore, this distance will have
to be large enough so that the moving vane does not collide with the
interferometer. Finally, the experiment will have to be carried out in a
high vacuum, so that the vane is not subject to random impacts from
atoms and molecules in the environment. Under these conditions,
motion of the vane will be essentially undamped, so that a method of
resetting it to zero after every impact may also have to be devised.

If, as described above, the atom interacts with the vane as a wave
and not as a particle, one may say that it interacts with the vane
\emph{as a whole}.\footnote{Chapter 7 of the book \emph{Particle
Metaphysics: A Critical Account of Subatomic Reality} by Brigitte
Falkenberg \cite{BF2007} is devoted to `Wave-particle duality'. The
author does not seem to have considered the possibility mentioned in
the above paragraph.}

Put differently, the interactions of a microscopic object with a
macroscopic one may be of two kinds.  The first kind would consist of
interactions of the microscopic object with atomic and sub-atomic
constituents of the the macroscopic object -- the vane, in the
experiment suggested. The second kind would consist of interactions
of the same microscopic object with the macroscopic object as a
whole. If interactions of the first kind are dominant, the vane will
see the incident object as a particle. It will see the incident
object as a wave only if interactions of the first kind are weak,
qualitatively speaking, by comparison with those of the second kind.
This condition may be difficult to meet if the microscopic object is
an elementary particle such as an electron, proton, neutron or photon
which interacts quite strongly with other elementary particles.
However, if it is an atom of a noble gas (e.g., He$^4$) with a long
de Broglie wavelength (without much penetrating power), the chances
of success may improve quite considerably. As the experiment could
not have been suggested before the advent of atom intereferometry, it
is fitting to refer the interested reader to a recent detailed survey
of the subject \cite{CSP2009}.

The existence of interactions of the second kind may be regarded as
aspect of quantum nonlocality in nonrelativistic physics.  

If the experiment succeeds, and one can follow the motion of the vane
with sufficient precision, it may be possible to tell whether both
kinds of interactions are simultaneously at work. Observation of
photons behaving simultaneously as waves and particles have been
reported by Foster et al \cite{FX2000}. However, these authors appear
to have been motivated not by Feynman's comments, but by the
Bohr-Kramers-Slater attempt of 1924, before the advent of quantum
mechanics, to reconcile the wave and particle properties of radiation
\cite{BKS1924}. For details, the reader is referred to an article by
Carmichael
\cite{HC2000}, one of the authors of \cite{FX2000}.

 \vspace{\baselineskip}\noindent {\bf Acknowledgements} I would like
to thank Samir Bose, Richard Kerner, Hansj\"org Roos, Michael Revzen
and Geoffrey Sewell for reading and criticizing earlier versions of
this note.

\end{document}